\documentclass{article}

\usepackage{graphicx}    % For inserting images
\usepackage{float}       % For [H] placement
\usepackage{amsmath}     % Math support
\usepackage{amssymb}     % \mathbb etc.
\usepackage{hyperref}    % Clickable refs & links
\usepackage{indentfirst} % Indent first paragraph
\usepackage{booktabs}    % Better tables
\usepackage{caption}
\usepackage{subcaption}
\usepackage{array}
\usepackage{geometry}
\usepackage{xcolor}
\usepackage{tabularx}
\usepackage{placeins}    % Float barriers

\newtheorem{proposition}{Proposition}

\title{Forecasting Liquidity Withdrawal with Machine Learning Models}

\author{%
  Haochuan (Kevin) Wang\\
  Massachusetts Institute of Technology\\
  \texttt{hcw@mit.edu}
}

\date{Aug 2025}

\begin{document}

\maketitle

\begin{abstract}
\noindent
Liquidity withdrawal is a critical indicator of market fragility. We develop and test a framework for forecasting liquidity withdrawal at the individual-stock level, spanning liquidity tiers from mid-cap names to highly liquid large-cap tickers, and evaluate the relative performance of competing model classes in predicting short-horizon order book stress. We propose the Liquidity Withdrawal Index (LWI)---defined as the ratio of order cancellations to the sum of smoothed standing depth and new additions at the best quotes---as a non-negative, scale-free, and interpretable measure of transient liquidity removal.

Using one hour of Nasdaq market-by-order (MBO) data (July 30, 2025, 14:00--15:00 ET) for seven symbols, we compare linear benchmarks (AR, HAR) and non-linear tree ensembles (XGBoost) across horizons ranging from 250\,ms to 5\,s. Predictability is horizon-dependent in our sample: forecasts at 250\,ms are noise-dominated; XGBoost delivers the best out-of-sample accuracy at 1\,s and 5\,s; and HAR is competitive, and best for most symbols, at 2\,s.

We discipline these findings in two ways. Theoretically, a linear AR(1)-plus-noise benchmark yields a closed-form upper bound on the $R^2$ attainable for aggregated targets, separating mechanical gains from target smoothing from genuine multi-scale predictability. Empirically, a fully reproducible end-to-end replication on public LOBSTER data (AAPL, June 21, 2012) with persistence baselines and Diebold--Mariano tests shows that the horizon structure and the value of multi-scale predictors generalize, while the level of predictability and the ranking of model classes are regime-dependent: in that wide-spread, sparse-book regime the top-of-book LWI is indistinguishable from short-memory noise and regularized linear models match or beat the tree ensemble. Forecastability is governed by the density and persistence of cancellation activity relative to noise, which our benchmark makes measurable ex ante.
\end{abstract}

\section{Introduction}
Displayed liquidity in modern equity markets can vanish within seconds as market makers cancel quotes and widen spreads, especially around announcements or bursts of asymmetric order placement. Anticipating such withdrawal even a few seconds ahead is valuable for execution, market making, and risk management across participants. This paper develops a framework for \emph{forecasting intraday liquidity withdrawal} at horizons from 250\,ms to 5\,s, using Nasdaq ITCH market-by-order (MBO) data and a set of econometric and machine learning models.  

\subsection{Liquidity Withdrawal Index (LWI)}
We reconstruct the limit order book in real time, align order messages to a 250\,ms grid, and define the Liquidity Withdrawal Index (LWI) as the fraction of liquidity removed at the top of book:
\[
\mathrm{LWI}_t \;=\; \frac{\text{Cancels}_t}{\mathrm{MA}_{1\text{s}}(\text{DepthL1})_{t-1} + \max(\text{Adds}_t,\epsilon)} .
\]
The denominator combines prior top-of-book depth with new additions, stabilized by a short moving average and a positive floor to avoid blow-ups in thin-book regimes. The resulting index is non-negative and scale-free rather than strictly bounded: because the smoothed denominator lags the current bin, intense cancellation bursts against a falling book can push the index well above one (visible as spikes in Figure~\ref{fig:rklb_lwi}). This interpretable index provides a real-time measure of order book fragility, directly tied to observable order flow.

\subsection{Feature Construction}
From the MBO stream we construct features including spreads at multiple depths, top-of-book depth, order flow imbalance (OFI), queue imbalance (QI), rolling means and volatilities, and activity rates (adds/cancels per second). These features capture both the state and dynamics of the book. A cross-ticker feature screening procedure yields a \emph{consensus} set dominated by short-horizon LWI lags and rolling volatilities of spread, depth, and order flow.

\subsection{Model Classes}
We compare two families of models across horizons of 250\,ms, 1\,s, 2\,s, and 5\,s:  
(i) linear baselines, AR(5) and HAR, which provide interpretable and latency-friendly forecasts;  
(ii) non-linear tree ensembles (XGBoost), which capture threshold effects and feature interactions.  
Evaluation uses walk-forward cross-validation with an embargo to mitigate information leakage and mimic real-time deployment.

\subsection{Contributions and Positioning}
\begin{figure}[H]
    \centering
    \includegraphics[width=0.8\linewidth]{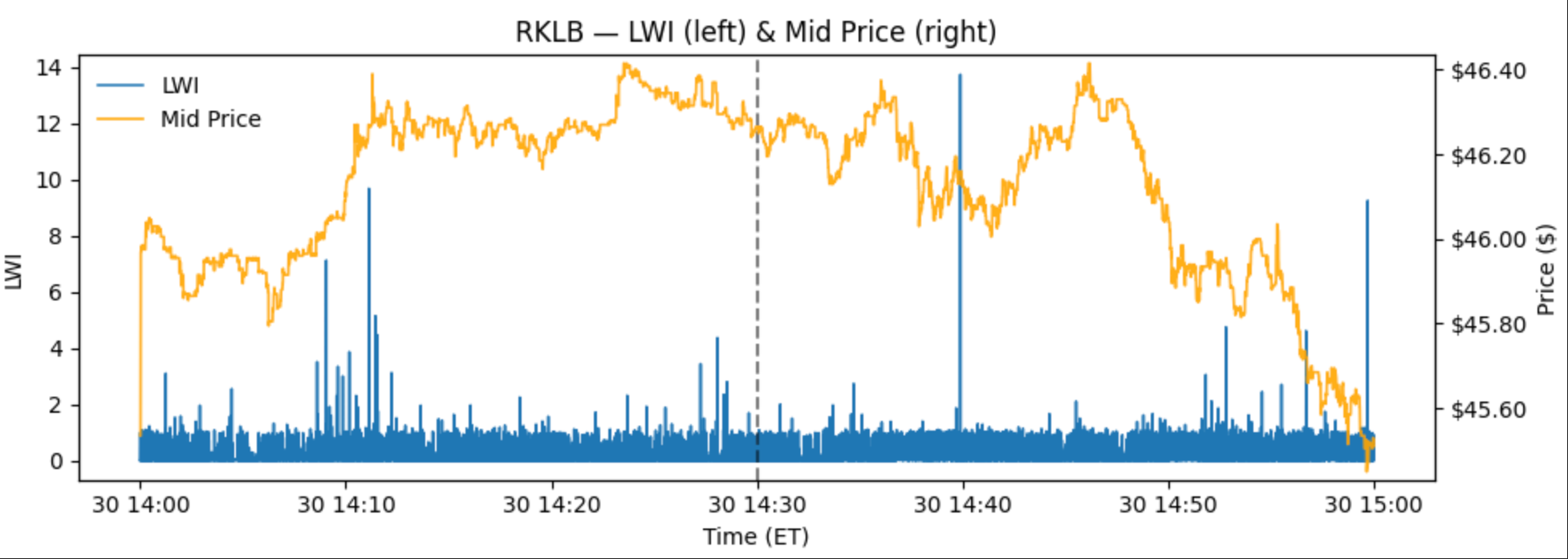}
    \caption{Liquidity Withdrawal Index (left axis) and mid price (right axis) for RKLB on July 30, 2025, 14:00--15:00 ET.}
    \label{fig:rklb_lwi}
\end{figure}

We make three contributions. First, we contribute a reproducible pipeline from raw MBO events to a stabilized LWI target and feature panel, and benchmark linear versus tree-based models under realistic evaluation. In our one-hour, seven-symbol sample, results show horizon-dependent structure: 250\,ms is noise-dominated; XGBoost delivers the best out-of-sample accuracy at 1\,s and 5\,s; and HAR is competitive, and best for most symbols, at the 2\,s horizon. Second, we derive a closed-form aggregation bound from a linear AR(1)-plus-noise benchmark (Section~\ref{sec:theory}) that separates mechanical $R^2$ gains from target smoothing from genuine multi-scale predictability, and whose parameters are identifiable from two autocovariances of any LWI sample. Third, we replicate the entire pipeline end-to-end on fully public LOBSTER data with persistence baselines and Diebold--Mariano tests (Section~\ref{sec:robustness}), showing which findings are structural and which are regime-specific. Unlike most LOB prediction studies that focus on mid-price direction, we target \emph{liquidity withdrawal}, aligning the analysis with operational priorities in execution and surveillance. Figure~\ref{fig:rklb_lwi} illustrates clustered LWI spikes for RKLB on July 30, 2025---around 14:10 ET, coinciding with an abrupt upward price gap, and again near 14:40 ET, where the largest spike accompanies a sharp price decline. Anticipating such withdrawal would allow market makers to adjust spreads or inventories proactively, reducing adverse selection risk.

\section{Literature Review}
\subsection{Liquidity, resiliency, and event-time dynamics}
Market quality is often summarized by spread, depth, and resiliency \cite{ohara1995microstructure,foucault2013liquidity}. Spread (bid--ask tightness) and depth (available volume) are widely studied, while resiliency---the speed of liquidity recovery after shocks---captures how quickly the book replenishes \cite{foucault2013liquidity}. Event studies show that liquidity conditions deteriorate around scheduled macroeconomic announcements: quoted depth falls and spreads widen around release times before normalizing \cite{scholtus2014news,hautsch2011news}, and in stress episodes such as the 2010 Flash Crash displayed liquidity can evaporate across venues within minutes \cite{menkveld2019flash}. Our Liquidity Withdrawal Index (LWI) formalizes this transient withdrawal phase at sub-second resolution, providing an interpretable, scale-free measure of book fragility that can be forecast and monitored in real time. It complements existing early-warning liquidity metrics such as order-flow toxicity (VPIN) \cite{easley2012vpin}: whereas VPIN infers adverse-selection risk from volume-bucketed trade imbalance, the LWI measures realized cancellation intensity at the best quotes directly.

\subsection{Working with MBO data in a low SNR regime}
Market-by-order (MBO) feeds provide granular detail on order flow (adds, cancels, replacements) but are noisy: bursts, flickering quotes, and irregular event timing. Effective engineering requires: (i) \emph{clocking} irregular events to a uniform grid, where the choice of sampling frequency trades off information against microstructure noise \cite{aitsahalia2005noise}; (ii) \emph{stabilizing} denominators to prevent ratio blow-ups in thin books; and (iii) \emph{filtering} via rolling means and variances to boost signal-to-noise, a standard step in feature construction for order-book prediction \cite{kercheval2015svm,ntakaris2018benchmark}. In line with this literature, our pipeline uses a 250\,ms grid with a warm start, denominator flooring, and rolling statistics to suppress flicker and heteroskedastic bursts before modeling.

\subsection{Model selection}
\label{sec:model_selection}

We evaluate two families of models, chosen to balance interpretability, predictive power, and latency.

\paragraph{Linear baselines: AR and HAR.}
Autoregressive (AR($p$)) models are simple and latency-friendly. We use AR(5), which provides a one-second lookback to capture short-term linear dependence in liquidity withdrawal. The Heterogeneous Autoregressive (HAR) model extends AR by including aggregates computed at multiple time scales \cite{corsi2009har}; the specific windows used in our implementation are given in Section~\ref{sec:results_protocol}. In our sample, HAR improves performance at the intermediate 2\,s horizon, but at the longer 5\,s horizon it is overtaken as non-linear dependencies grow more important.

\paragraph{Tree ensembles and threshold effects.}
Gradient-boosted decision trees (XGBoost) capture conditional thresholds and feature interactions efficiently. Prior work finds that non-linear methods---tree ensembles and neural networks---often outperform linear benchmarks in financial prediction, both in the cross-section of returns \cite{gu2020ml} and in limit order book forecasting \cite{sirignano2019universal,ntakaris2018benchmark}. In our July 2025 sample, XGBoost substantially outperforms AR/HAR at the 1\,s and 5\,s horizons; our public-data robustness study (Section~\ref{sec:robustness}) shows, however, that this ranking is regime-dependent---in a sparse-book regime, a regularized linear model on the same features performs as well or better.

\subsection{Positioning and implications}
Unlike much of the LOB prediction literature focused on mid-price direction \cite{ntakaris2018benchmark,sirignano2019universal}, we target \emph{liquidity withdrawal} directly, through a ratio observable from cancels and adds. This target is interpretable and operational: predictions map to a tangible phenomenon that market participants can act upon. In our sample, boosted trees deliver the best accuracy at the 1\,s and 5\,s horizons, HAR is the strongest model for most symbols at 2\,s, and no model delivers robust predictability at 250\,ms. These findings translate into actionable design principles for execution and surveillance systems, where early warning of transient liquidity stress is most valuable.

\section{Data}
We use Market-by-Order (MBO) data for U.S. equities from Databento.\footnote{\url{https://databento.com/portal/catalog/us-equities}}  
MBO provides the full order flow --- every add, cancel, replace, and trade message from the NASDAQ matching engine --- at nanosecond resolution. For tractability, our study focuses on July 30, 2025, 14:00--15:00 ET, covering seven actively traded symbols spanning two liquidity tiers: three highly liquid large-cap names (AAPL, NVDA, TSLA) and four actively traded mid-cap names (HIMS, NBIS, RKLB, SNAP). After aligning messages to a 250\,ms Eastern Time grid and reconstructing the top of book, we obtain a panel of 14{,}400 observations per symbol (four bins per second over one hour). This resampling captures sub-second dynamics while taming the burstiness of raw event time. We emphasize that the narrow one-hour, single-day window keeps the study tractable but limits external validity; extensions to longer samples are discussed in the conclusion.

From the resampled book we construct both the target, the Liquidity Withdrawal Index (LWI), and a panel of microstructure features: spread and depth at level~1, order and queue imbalances (OFI, QI), rolling means and volatilities of mid-price returns, and add/cancel rates. These engineered features provide interpretable summaries of state, flow, and short-term volatility. A full list of raw MBO fields is provided in Appendix~\ref{app:datafields}; the derived features are described in Section~\ref{sec:methods}.

\section{Methods}
\label{sec:methods}

Our methodology follows a pipeline from raw MBO data to engineered features, then to model estimation and evaluation across multiple horizons. The design emphasizes predictive accuracy while maintaining robustness and low latency in high-noise, high-frequency settings.

\subsection{Data Preparation}
We reconstruct the limit order book from Databento Market-by-Order (MBO) events and resample all messages onto a uniform 250\,ms Eastern Time (ET) grid. This step aligns the inherently irregular event stream and avoids aliasing ultrafast bursts. It also facilitates modeling by producing equally spaced time-series suitable for both linear and non-linear estimators.

From the reconstructed book, we compute a panel of features commonly used in market microstructure research: bid--ask spread, top-of-book depth, order flow imbalance (OFI) \cite{cont2014ofi}, queue imbalance (QI), rolling means and standard deviations of LWI and QI, realized variance proxies, and activity rates (adds/cancels). Feature selection combines mutual information, gradient-boosted tree importance, and LASSO, yielding a consensus set dominated by short-horizon LWI lags, moving averages, and volatility measures.

\subsection{Feature Selection Framework}
Because high-frequency order flow is noisy and collinear, feature selection must balance stability with interpretability. We therefore evaluate each feature’s informativeness using three complementary methods:

\paragraph{Mutual Information (MI).}
MI measures general dependence between feature $X$ and target $Y$, capturing both linear and non-linear associations. For LWI, MI identifies which features carry the strongest predictive content regardless of functional form.

\paragraph{XGBoost Feature Importance.}
Tree ensembles rank features by their marginal contribution to predictive accuracy. XGBoost is particularly well suited to detect threshold effects such as ``if queue imbalance is extreme and depth collapses, withdrawal rises.’’

\paragraph{LASSO Regression.}
The $\ell_1$ penalty of LASSO enforces sparsity, shrinking weak predictors to zero and highlighting features with stable linear contributions. A feature that consistently ranks highly across all three approaches is considered robust and included in the consensus panel.

\subsection{Consensus Across Tickers}
We apply the selection framework at the 1-second horizon across four of the seven study symbols (HIMS, NBIS, RKLB, SNAP) and aggregate the results; the resulting consensus feature set is then used for all seven symbols in the evaluation of Section~\ref{sec:results_protocol}. Table~\ref{tab:consensus} reports, for each feature, the number of symbols in which it appears among the top predictors, its mean rank, and total method hits. Consensus features are those appearing in at least 60\% of tickers.

\begin{table}[H]
\centering
\small
\begin{tabular}{lcccc}
\toprule
Feature & $n_{\text{symbols}}$ & Mean Best Rank & Method Hits & Consensus \\
\midrule
LWI\_ma1s      & 4 & 1.00 & 12 & Yes \\
LWI\_lag1      & 4 & 1.00 &  8 & Yes \\
LWI\_sd1s      & 4 & 2.00 & 10 & Yes \\
dLWI\_1s       & 4 & 2.50 & 12 & Yes \\
LWI\_lag2      & 4 & 2.75 &  8 & Yes \\
adds\_rate1s   & 4 & 5.00 &  8 & Yes \\
canc\_rate1s   & 4 & 5.75 &  6 & Yes \\
QI\_lag1s      & 4 & 5.75 &  4 & Yes \\
depth\_L1\_lag1s & 4 & 5.75 & 4 & Yes \\
depth\_L1\_lag4  & 4 & 6.00 & 4 & Yes \\
QI\_sd1s       & 4 & 6.25 &  6 & Yes \\
LWI\_ma10s     & 4 & 6.75 &  5 & Yes \\
QI\_lag4       & 4 & 6.75 &  4 & Yes \\
LWI\_ma2s      & 4 & 8.00 &  6 & Yes \\
LWI\_sd2s      & 3 & 7.00 &  4 & Yes \\
spread\_sd1s   & 3 & 8.00 &  5 & Yes \\
\bottomrule
\end{tabular}
\caption{Consensus feature selection across four symbols at the 1-second horizon.}
\label{tab:consensus}
\end{table}

\subsection{Stationarity and Persistence}
We test whether the Liquidity Withdrawal Index (LWI) is stationary using the Augmented Dickey–Fuller (ADF) test. Across all four tickers, the ADF statistics are strongly negative with p-values near zero (Table~\ref{tab:adf}), rejecting the unit root null. Thus, LWI is stationary, making it valid for autoregressive modeling without differencing. Autocorrelation plots reveal significant dependence at the first one or two lags, followed by rapid decay. This pattern is consistent with a stationary but short-memory process: shocks to liquidity withdrawal are temporary but persist briefly, reinforcing the use of short lags as predictive signals.

\begin{table}[H]
\centering
\begin{tabular}{lrr}
\toprule
\textbf{Symbol} & \textbf{ADF statistic} & \textbf{p-value} \\
\midrule
HIMS & -17.97 & $2.79 \times 10^{-30}$ \\
NBIS & -13.61 & $1.88 \times 10^{-25}$ \\
RKLB & -11.89 & $6.03 \times 10^{-22}$ \\
SNAP & -12.71 & $1.01 \times 10^{-23}$ \\
\bottomrule
\end{tabular}
\caption{ADF test results for LWI stationarity.}
\label{tab:adf}
\end{table}

\begin{figure}[H]
    \centering
    \includegraphics[width=0.5\linewidth]{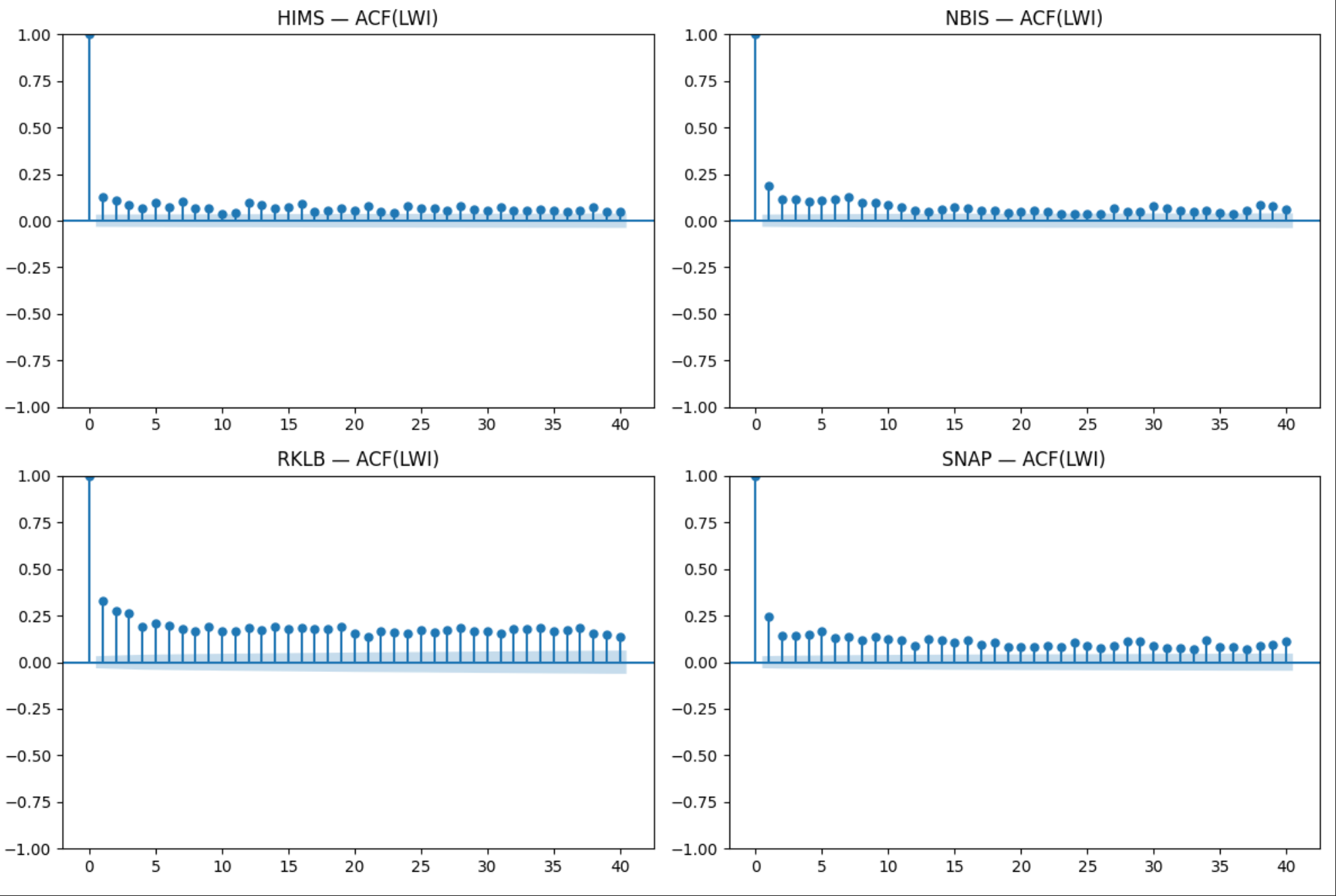}
    \caption{ACF and PACF of LWI across four tickers, showing short-memory dynamics.}
    \label{fig:acf}
\end{figure}

\subsection{A Linear Benchmark: Noise, Persistence, and Target Aggregation}
\label{sec:theory}
Because the forecast target at horizon $k$ is the \emph{average} of the next $k$ bins, mechanical effects of aggregation must be separated from genuine predictability. We use the simplest linear state-space model that captures both persistence and measurement noise:
\begin{equation}
y_t = s_t + u_t, \qquad s_t = \varphi\, s_{t-1} + \eta_t, \qquad 0<\varphi<1,
\label{eq:arnoise}
\end{equation}
where $y_t$ is the observed LWI, $s_t$ is a stationary AR(1) cancellation-intensity signal with variance $\sigma_s^2$, and $u_t$ is white measurement noise with variance $\sigma_u^2$, independent of the signal.

\begin{proposition}[Aggregation bound]\label{prop:aggregation}
Let $\bar y_{t,k} = k^{-1}\sum_{j=1}^{k} y_{t+j}$. Under model \eqref{eq:arnoise}, the population $R^2$ of the optimal forecast of $\bar y_{t,k}$ based on time-$t$ information is
\[
\bar R^2(k) \;=\; \frac{B_k}{A_k + \sigma_u^2/k}, \qquad
B_k = \frac{\sigma_s^2\,\varphi^2\,(1-\varphi^k)^2}{k^2\,(1-\varphi)^2}, \qquad
A_k = \frac{\sigma_s^2}{k^2}\Big(k + 2\sum_{d=1}^{k-1}(k-d)\,\varphi^d\Big),
\]
and no forecast based on the history of $y$ alone attains a higher population $R^2$. (Proof in Appendix~B.)
\end{proposition}

Two properties of $\bar R^2(k)$ matter for interpreting our results. First, aggregation shrinks the noise contribution $\sigma_u^2/k$ at rate $1/k$, so measured $R^2$ can \emph{rise} with the horizon even though no additional information is being exploited: high $R^2$ on aggregated targets partly reflects a smoother target. Second, once $k$ exceeds the signal's decorrelation time, the numerator decays faster than the denominator and $\bar R^2(k)$ \emph{falls}; under a single short-memory component, long-horizon aggregated targets are less predictable, not more. The parameters are identified from the first two autocovariances of the observed series ($\varphi = \gamma_2/\gamma_1$, $\sigma_s^2 = \gamma_1/\varphi$, $\sigma_u^2 = \gamma_0 - \sigma_s^2$), so the bound can be computed on any sample and confronted with realized out-of-sample $R^2$. Realized values \emph{above} the bound for forecasts that use only the LWI's own history reject the single-component model and indicate additional, more slowly varying state---precisely the multi-scale structure that HAR encodes. We apply this diagnostic to public data in Section~\ref{sec:robustness}.

\subsection{Evaluation Protocol}
We apply walk-forward cross-validation with an embargo period to prevent information leakage. Each split trains on earlier intervals and tests on later ones, emulating live deployment. Models are evaluated using out-of-sample $R^2$ across folds and horizons. Because the $k$-step target is an average over the next $k$ bins, temporal aggregation smooths the target and mechanically raises attainable $R^2$ at longer horizons; cross-horizon comparisons should therefore be read qualitatively rather than as like-for-like. Reporting horizon-specific performance reveals which models dominate at different timescales: in our sample, XGBoost is strongest at 1\,s and 5\,s, HAR at 2\,s, with minimal predictability at 250\,ms where noise dominates.

\subsection{Summary of Design Choices}
Our pipeline integrates production constraints with research aims:  
(i) \textbf{Uniform resampling and filtering} stabilize noisy MBO streams.  
(ii) \textbf{Consensus features} ensure robustness across tickers.  
(iii) \textbf{Linear models} capture short-memory dependencies in LWI.  
(iv) \textbf{Boosted trees} provide a flexible non-linear benchmark.  
Together, these design choices allow us to characterize when linear vs.~non-linear structure dominates and how feature-engineered predictors translate into operationally relevant liquidity withdrawal forecasts.

\section{Results}
\subsection{Evaluation Protocol}
\label{sec:results_protocol}
We use a five-fold expanding-window walk-forward split with an embargo buffer to prevent look-ahead bias. In each fold, models are trained on all prior data and tested on the next out-of-sample block, mirroring live deployment.  

Targets are defined as the mean Liquidity Withdrawal Index (LWI) over the next $k$ bins, where $k=1$ for 250\,ms and $k=20$ for 5\,s. Thus, 250\,ms corresponds to a one-step forecast, while 5\,s is a many-step forecast. Benchmarks follow our implementation:  
\begin{itemize}
    \item {AR(5)}: short-memory autoregression with lags up to 1\,s.  
    \item {HAR}: heterogeneous autoregression with aggregates at 250\,ms, 2\,s, and 10\,s.  
    \item {XGB}: boosted trees on lagged and rolling engineered features.  
\end{itemize}

\subsection{Performance Across Models and Horizons}
Table~\ref{tab:all_r2} summarizes mean out-of-sample $R^2$ by symbol, model, and horizon. Patterns are clearly horizon-dependent:
\begin{itemize}
    \item At 250\,ms, out-of-sample $R^2$ is near zero (or negative) for all models: the one-step target is noise-dominated.
    \item At 1\,s, XGB is the best model for all seven symbols. HAR improves on AR for only three of the seven symbols (AAPL, NVDA, TSLA), so multi-scale linear aggregates alone do not reliably help at this horizon.
    \item At 2\,s, HAR improves on AR for every symbol and is the best model for five of seven; XGB leads for AAPL and NVDA by small margins.
    \item At 5\,s, XGB dominates for all seven symbols, with $R^2$ above 0.90 for six of seven. Part of this level reflects the smoother aggregated target (a 20-bin average) rather than model skill alone. AR(5) deteriorates sharply at this horizon (negative $R^2$ for four symbols), indicating that projecting short 250\,ms lags onto a multi-second aggregated target is misspecified.
\end{itemize}

\begin{table}[H]
\centering
\small
\begin{tabular}{llcccc}
\toprule
\textbf{Symbol} & \textbf{Model} & \textbf{250\,ms} & \textbf{1\,s} & \textbf{2\,s} & \textbf{5\,s} \\
\midrule
AAPL & AR(5) & -0.146 & 0.468 & 0.632 & -0.684 \\
     & HAR   & -0.008 & 0.527 & 0.850 & 0.727 \\
     & XGB   & \textbf{0.006} & \textbf{0.786} & \textbf{0.862} & \textbf{0.950} \\
\midrule
NVDA & AR(5) & -0.294 & 0.462 & 0.649 & -1.245 \\
     & HAR   & \textbf{0.053} & 0.640 & 0.904 & 0.799 \\
     & XGB   & 0.049 & \textbf{0.843} & \textbf{0.913} & \textbf{0.963} \\
\midrule
TSLA & AR(5) & -0.098 & 0.499 & 0.658 & -0.270 \\
     & HAR   & -0.043 & 0.519 & \textbf{0.850} & 0.735 \\
     & XGB   & -0.225 & \textbf{0.706} & 0.822 & \textbf{0.936} \\
\midrule
HIMS & AR(5) & -0.023 & 0.533 & 0.681 & 0.061 \\
     & HAR   & -0.006 & 0.466 & \textbf{0.821} & 0.653 \\
     & XGB   & -0.053 & \textbf{0.677} & 0.777 & \textbf{0.892} \\
\midrule
NBIS & AR(5) & 0.002 & 0.587 & 0.743 & 0.294 \\
     & HAR   & \textbf{0.034} & 0.546 & \textbf{0.866} & 0.731 \\
     & XGB   & -0.209 & \textbf{0.689} & 0.838 & \textbf{0.938} \\
\midrule
RKLB & AR(5) & -0.060 & 0.544 & 0.715 & -0.053 \\
     & HAR   & \textbf{0.006} & 0.539 & \textbf{0.867} & 0.723 \\
     & XGB   & -0.057 & \textbf{0.776} & 0.859 & \textbf{0.930} \\
\midrule
SNAP & AR(5) & -0.022 & 0.545 & 0.699 & 0.147 \\
     & HAR   & \textbf{0.012} & 0.497 & \textbf{0.837} & 0.703 \\
     & XGB   & -0.142 & \textbf{0.761} & 0.849 & \textbf{0.931} \\
\bottomrule
\end{tabular}
\caption{Out-of-sample $R^2$ across models (AR(5), HAR, XGB) and horizons (250\,ms–5\,s). Bold values indicate the best model per symbol-horizon.}
\label{tab:all_r2}
\end{table}
  % <- external table file, otherwise paste directly

\subsection{Boosted Trees Diagnostics}
Gradient-boosted trees deliver the strongest performance at the 1\,s and 5\,s horizons. At 1\,s, XGB outperforms AR/HAR for every symbol but underpredicts sharp cancellation spikes, producing right-skewed residuals. At 5\,s, temporal aggregation smooths noise and residuals become symmetric and centered, indicating better calibration.  

Two adjustments improve short-horizon accuracy:  
(i) spike-aware losses (Huber, quantile, or Box--Cox targets) to reduce underprediction of extreme withdrawals;  
(ii) stress-sensitive features (depth drawdowns, cancel/replace bursts) to capture imminent liquidity shocks.  

\begin{figure}[H]
    \centering
    \includegraphics[width=1\linewidth]{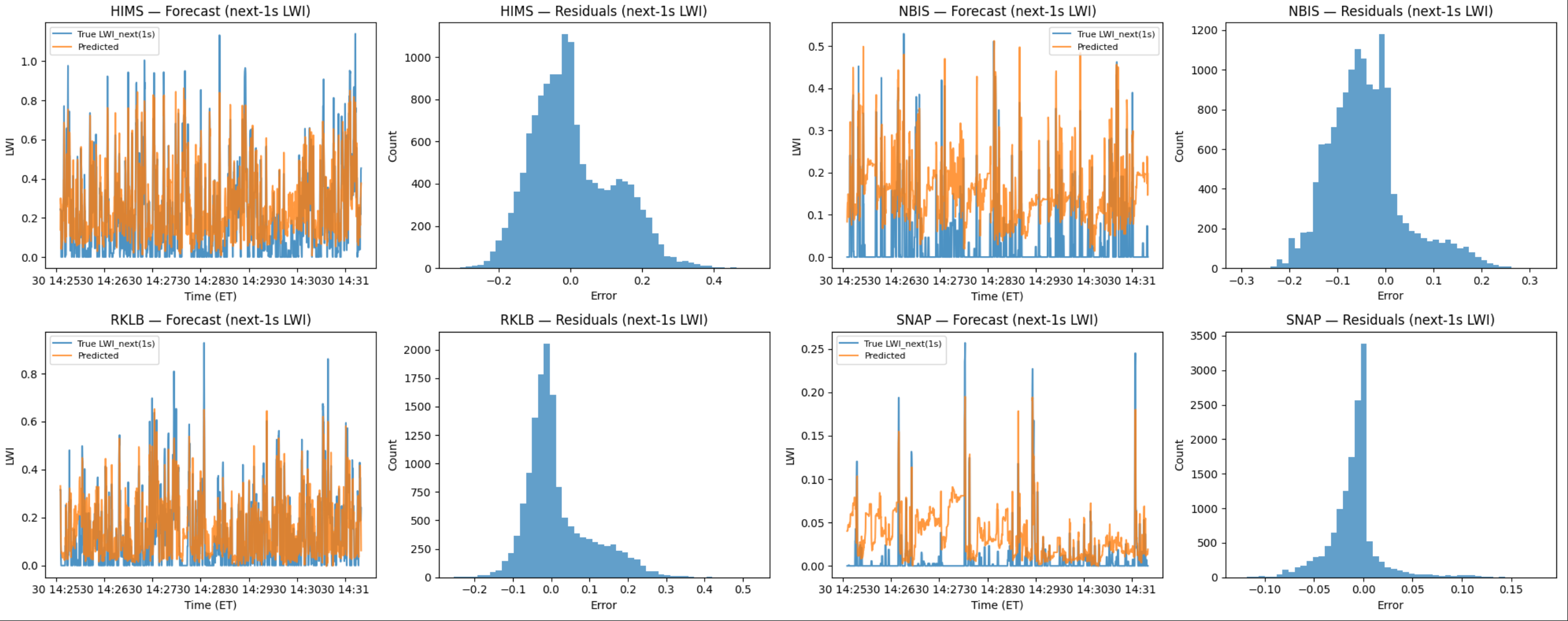}
    \caption{\textbf{XGB one-second LWI forecast.}}
\end{figure}

\begin{figure}[H]
    \centering
    \includegraphics[width=1\linewidth]{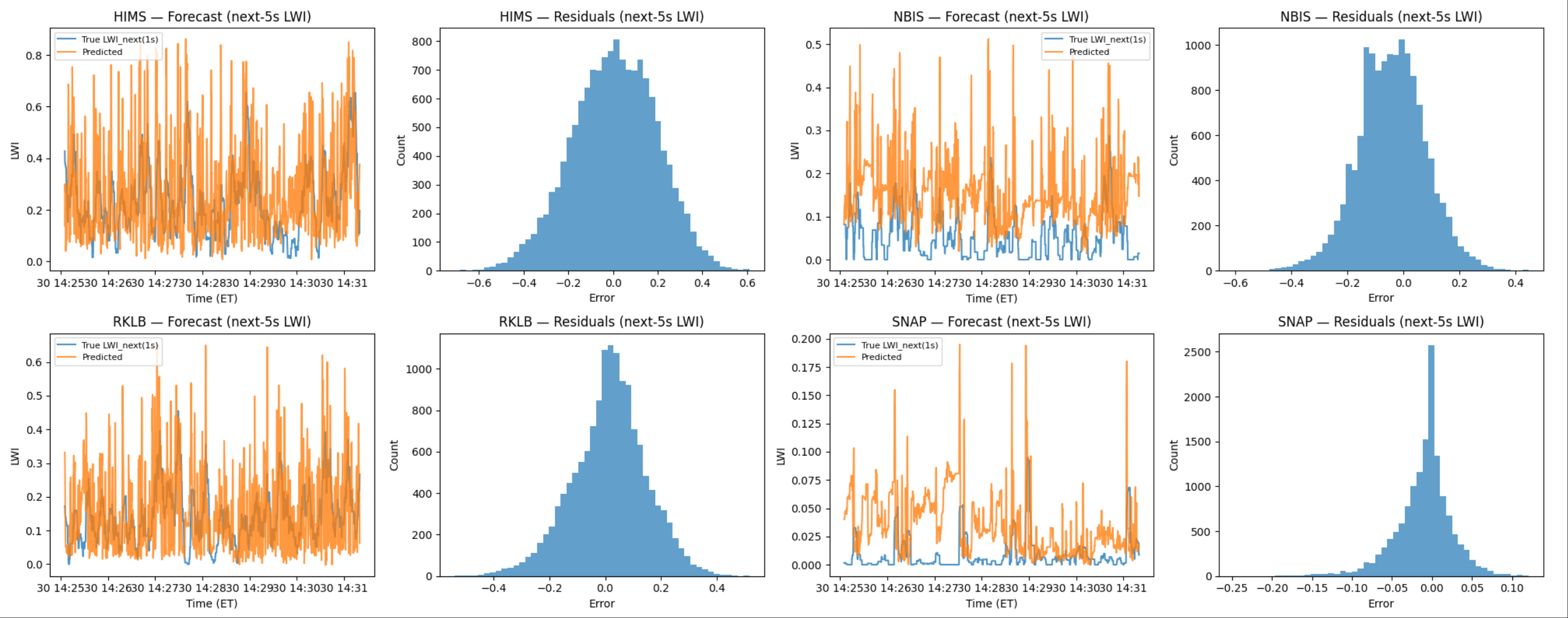}
    \caption{\textbf{XGB five-second LWI forecast.} 
    Temporal aggregation yields higher $R^2$ and near-symmetric residuals, indicating improved calibration.}
\end{figure}

\begin{figure}[H]
    \centering
    \includegraphics[width=0.8\linewidth]{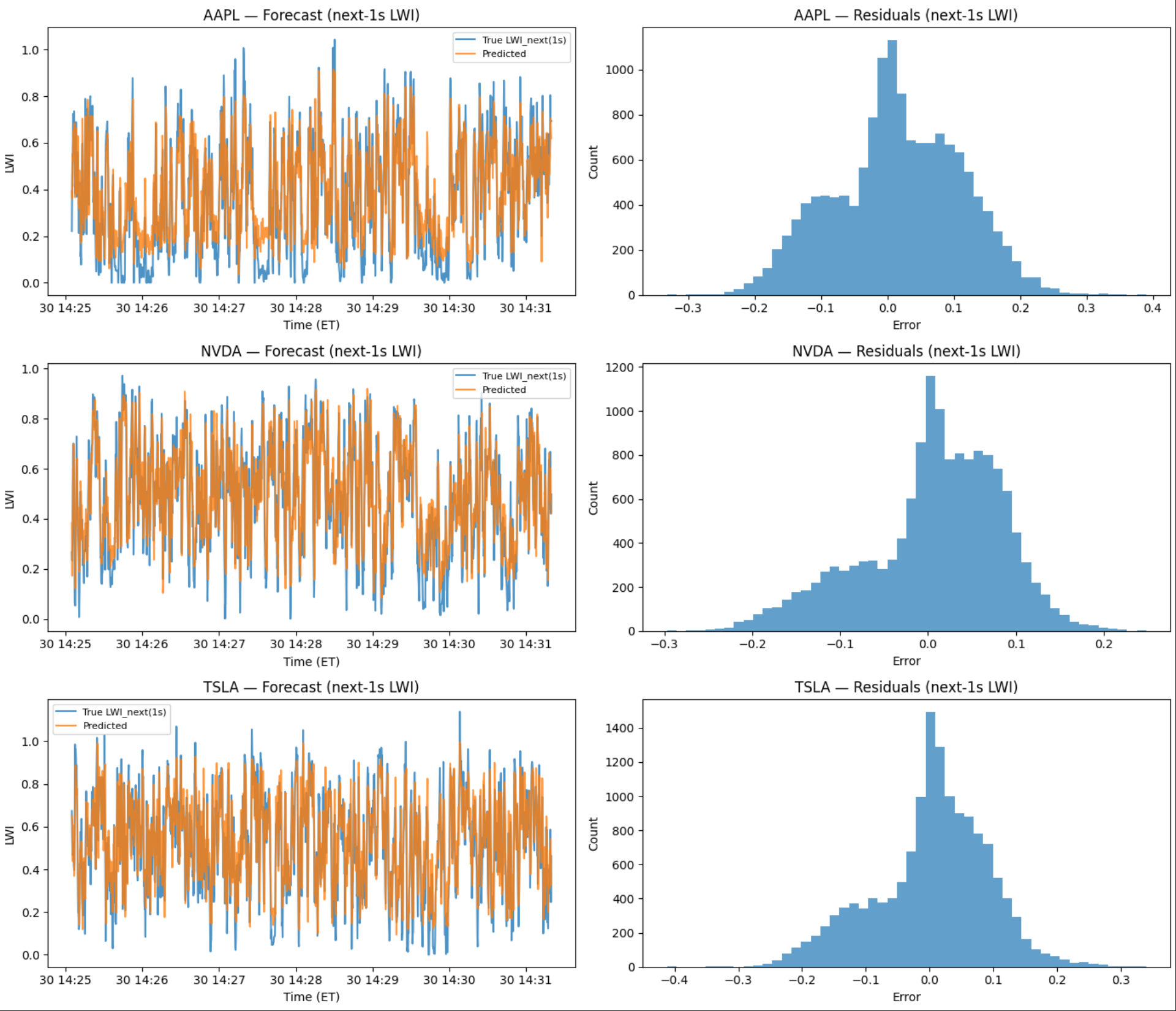}
    \caption{\textbf{XGB one-second LWI forecasts.} 
    For highly liquid tickers such as AAPL, NVDA, and TSLA, residuals are less skewed even at the 1\,s horizon. 
    This suggests fundamental differences in risk dynamics across liquidity tiers, motivating specialized models for distinct liquidity clusters rather than a uniform approach.}
\end{figure}

\FloatBarrier

\section{Robustness on Public Data}
\label{sec:robustness}
The results of Section~5 come from one hour of commercial data. To assess external validity and make part of the study fully reproducible, we replicate the entire pipeline end-to-end on a \emph{public} sample: the official LOBSTER sample file for AAPL on June~21, 2012 (reconstructed from Nasdaq Historical TotalView-ITCH), containing 400{,}391 messages with ten levels of book depth over the full trading day.\footnote{Provenance, retrieval parameters, and SHA-256 file hashes are recorded in \texttt{DATA\_ACCESS.md} in the replication repository; the derived 250\,ms panel and all code to rebuild it are committed there. The commercial July 2025 sample of Section~3 cannot be redistributed.} This sample differs from Section~3 by venue vintage (2012 vs.\ 2025), liquidity regime, and tick regime: AAPL traded near \$585 with an average quoted spread of about 15 cents, so the top of book is comparatively sparse. Precisely because the regime differs, the exercise reveals which findings are structural and which are sample-specific.

\subsection{Protocol}
The pipeline is identical to Sections~4--5: messages are aligned to a 250\,ms grid (93{,}600 bins, 09:30--16:00 ET), the LWI and the feature panel are constructed as before, and models are evaluated with the same five-fold expanding walk-forward split with a 10\,s embargo. We evaluate on 10:00--15:30 ET (79{,}200 bins) to avoid open and close effects. Three additions address evaluation concerns raised in Section~\ref{sec:theory}: (i) a \emph{persistence} baseline that forecasts the next-$k$-bin mean with the trailing $k$-bin mean; (ii) a \emph{Ridge} regression on the full feature set, to separate the value of the covariates from the value of non-linearity; and (iii) Diebold--Mariano (DM) tests on squared-error differentials with Newey--West variances (lag $k+4$, accommodating the overlap of aggregated targets). We compute two LWI variants: the \emph{top-of-book} definition of Section~1, and a \emph{book-wide} variant that counts all visible additions and cancellations over ten levels against total visible depth.

\begin{table}[H]
\centering
\small
\begin{tabular}{lrrrrrrr}
\toprule
\textbf{LWI variant} & \textbf{Mean} & \textbf{SD} & \textbf{Skew} & \textbf{\% zero} & \textbf{\% $>1$} & \textbf{ACF(1)} & \textbf{ACF(40)} \\
\midrule
Top-of-book & 0.051 & 0.241 & 95.9 & 85.5 & 0.39 & 0.055 & 0.011 \\
Book-wide & 0.040 & 0.082 & 3.5 & 57.7 & 0.00 & 0.291 & 0.045 \\
\bottomrule
\end{tabular}
\caption{Distribution of the 250\,ms LWI for AAPL on 2012-06-21 (10:00--15:30 ET). The top-of-book variant is zero in most bins and spikes far above one in this wide-spread, sparse-book regime; the book-wide variant remains below one in this sample.}
\label{tab:lobster_stats}
\end{table}

\subsection{Results}
Table~\ref{tab:lobster_stats} shows that the two variants inhabit different regimes. Top-of-book cancellations are rare in this wide-spread stock: 85.5\% of 250\,ms bins contain none, the index is extremely right-skewed, and first-order autocorrelation is only 0.055. The book-wide index is far less sparse (57.7\% zero bins), bounded observations remain below one, and autocorrelation is substantial (0.29 at one lag).

\begin{table}[H]
\centering
\small
\begin{tabular}{lcccc}
\toprule
\textbf{Model} & \textbf{250\,ms} & \textbf{1\,s} & \textbf{2\,s} & \textbf{5\,s} \\
\midrule
\multicolumn{5}{l}{\emph{Panel A: top-of-book LWI (paper definition)}}\\
Persistence & -0.850 & -0.734 & -0.678 & -0.638 \\
AR(5) & 0.005 & 0.005 & -0.005 & -0.039 \\
HAR & 0.010 & 0.018 & 0.017 & 0.001 \\
Ridge & \textbf{0.037} & \textbf{0.053} & \textbf{0.048} & \textbf{0.028} \\
XGB & 0.029 & 0.038 & 0.024 & -0.026 \\
\midrule
\multicolumn{5}{l}{\emph{Panel B: book-wide LWI (all visible levels)}}\\
Persistence & -0.441 & -0.383 & -0.370 & -0.313 \\
AR(5) & 0.097 & 0.105 & 0.095 & 0.074 \\
HAR & 0.100 & 0.126 & 0.132 & 0.136 \\
Ridge & \textbf{0.114} & \textbf{0.143} & \textbf{0.150} & \textbf{0.153} \\
XGB & 0.087 & 0.092 & 0.083 & 0.078 \\
\bottomrule
\end{tabular}
\caption{Out-of-sample $R^2$ on LOBSTER AAPL (2012-06-21, 10:00--15:30 ET), mean across five expanding walk-forward folds with a 10\,s embargo. Persistence forecasts the next-$k$-bin mean with the trailing $k$-bin mean. Bold marks the best model per horizon within each panel.}
\label{tab:lobster_r2}
\end{table}

Table~\ref{tab:lobster_r2} reports out-of-sample $R^2$. Four findings stand out.
\begin{itemize}
  \item \textbf{Top-of-book LWI is essentially unforecastable in this regime.} The best model attains $R^2 = 0.053$ (Ridge, 1\,s); AR(5) and XGBoost turn negative at 5\,s. Predictability of the paper's headline magnitude is absent.
  \item \textbf{The book-wide variant is moderately forecastable, and $R^2$ \emph{rises} with the horizon} for the feature-based models (Ridge: 0.114 at 250\,ms to 0.153 at 5\,s; HAR: 0.100 to 0.136). The direction of the horizon dependence in Section~5 therefore replicates; its magnitude does not.
  \item \textbf{Non-linearity is not the source of the gains here.} Ridge on the same features matches or beats XGBoost at every horizon in both panels; the difference is statistically significant in Ridge's favor at 5\,s for the top-of-book variant (DM $=2.59$, $p=0.010$) and indistinguishable at 1--2\,s. Tree dominance is regime-dependent, not universal.
  \item \textbf{The trailing-mean persistence baseline is sharply negative} in both panels: with a bursty, zero-inflated target, recent averages are poor forecasts, and the relevant naive benchmark is the unconditional mean. All models beat persistence (DM $p<0.07$ at every horizon).
\end{itemize}
The definitional sensitivity checks (Table~\ref{tab:lobster_sens}) leave these conclusions unchanged: varying the depth-smoothing window (0.5\,s/1\,s/2\,s), the additions floor $\epsilon$, and unsmoothed depth moves top-of-book $R^2$ by at most a few thousandths, and the model ranking is unaffected.

\begin{table}[H]
\centering
\small
\begin{tabular}{lrrrr}
\toprule
 & \multicolumn{2}{c}{\textbf{1\,s}} & \multicolumn{2}{c}{\textbf{5\,s}} \\
\textbf{LWI definition variant} & HAR & XGB & HAR & XGB \\
\midrule
MA window 0.5\,s & 0.017 & 0.032 & -0.005 & -0.049 \\
MA window 2\,s & 0.016 & 0.028 & 0.006 & -0.017 \\
$\epsilon=10$ shares & 0.019 & 0.040 & 0.003 & -0.032 \\
unsmoothed depth & 0.014 & 0.022 & -0.012 & -0.066 \\
\bottomrule
\end{tabular}
\caption{Sensitivity of out-of-sample $R^2$ (mean across folds) to the top-of-book LWI construction: trailing-MA window of the depth denominator, additions floor $\epsilon$, and unsmoothed depth.}
\label{tab:lobster_sens}
\end{table}

\subsection{Confronting the aggregation bound}
Table~\ref{tab:lobster_theory} calibrates Proposition~\ref{prop:aggregation} on each variant by the method of moments and compares the single-component bound $\bar R^2(k)$ with realized performance. The estimated signal share $\sigma_s^2/(\sigma_s^2+\sigma_u^2)$ is 9.1\% for the top-of-book index ($\hat\varphi=0.61$) and 41.7\% for the book-wide index ($\hat\varphi=0.70$).

\begin{table}[H]
\centering
\small
\begin{tabular}{llrrr}
\toprule
\textbf{LWI variant} & \textbf{Horizon} & \textbf{Bound $\bar R^2(k)$} & \textbf{Univariate (AR/HAR)} & \textbf{Ridge (full features)} \\
\midrule
Top-of-book & 250ms & 0.034 & 0.010 & 0.037 \\
 & 1s & 0.037 & 0.018 & 0.053 \\
 & 2s & 0.022 & 0.017 & 0.048 \\
 & 5s & 0.009 & 0.001 & 0.028 \\
Book-wide & 250ms & 0.203 & 0.100 & 0.114 \\
 & 1s & 0.190 & 0.126 & 0.143 \\
 & 2s & 0.115 & 0.132$^{\dagger}$ & 0.150 \\
 & 5s & 0.043 & 0.136$^{\dagger}$ & 0.153 \\
\bottomrule
\end{tabular}
\caption{Calibrated single-component bound $\bar R^2(k)$ of Proposition~\ref{prop:aggregation} (method-of-moments parameters from the same sample) versus realized out-of-sample $R^2$. The bound constrains only forecasts built from the LWI's own history (AR/HAR); $^{\dagger}$ marks univariate performance above the bound, which rejects the single-component model and indicates multi-scale structure. Ridge uses order-book covariates and is not constrained by the bound.}
\label{tab:lobster_theory}
\end{table}

For the top-of-book variant, univariate forecasts (AR/HAR, functions of the LWI's own history) remain \emph{below} the bound at every horizon: the series is statistically indistinguishable from a single short-memory component plus noise, and only order-book covariates (Ridge) add modest information. For the book-wide variant, HAR \emph{exceeds} the bound at 2\,s and 5\,s---by a factor of three at 5\,s---which rejects the single-component model and establishes genuinely multi-scale cancellation-intensity dynamics: slow components dominate long-horizon predictability, exactly the structure HAR's multi-scale aggregates encode. This also disciplines the interpretation of high $R^2$ on aggregated targets: under a single component the bound \emph{falls} with $k$ beyond the memory scale, so any $R^2$ that rises with the horizon is evidence of slow state variables, and its level should be judged against the calibrated bound rather than against zero.

\subsection{What generalizes, and what does not}
Combining Sections~5 and~6: the \emph{horizon-dependence} of LWI forecastability and the value of multi-scale predictors generalize across two venues thirteen years apart; the \emph{level} of predictability and the ranking of model classes do not. Forecastability is governed by the density and persistence of cancellation activity relative to measurement noise---the signal share of Section~\ref{sec:theory}---which varies with the liquidity and tick regime of the stock and with the LWI definition (top-of-book versus book-wide). We therefore recommend reporting the estimated signal share and the calibrated bound $\bar R^2(k)$ alongside any LWI forecasting results; for the July 2025 sample of Section~5 this is left for future work, as the underlying commercial data cannot be redistributed with this paper.

\FloatBarrier

\section{Conclusion}
We propose and evaluate a framework for forecasting liquidity withdrawal using Nasdaq MBO data and a Liquidity Withdrawal Index (LWI). In the July 2025 commercial sample, results across seven tickers show a horizon-dependent structure: forecasts are noise-dominated at 250\,ms; XGBoost delivers the best out-of-sample accuracy at 1\,s and 5\,s, reaching $R^2$ above 0.90 at 5\,s for six of seven symbols; and HAR is competitive, and best for most symbols, at 2\,s.

The fully reproducible public-data study of Section~\ref{sec:robustness} disciplines these findings. The horizon-dependence and the value of multi-scale predictors generalize: on 2012 AAPL data, book-wide cancellation intensity is forecastable with $R^2$ rising in the horizon, and HAR exceeds the calibrated single-component bound of Proposition~\ref{prop:aggregation}, confirming multi-scale structure. The headline magnitudes and the model ranking do not generalize: in that wide-spread, sparse-book regime the top-of-book LWI is statistically indistinguishable from short-memory noise, and a Ridge regression on the same features matches or beats the tree ensemble. Together the two samples suggest that LWI forecastability is governed by the density and persistence of cancellation activity relative to noise---a quantity that our linear benchmark makes measurable ex ante---rather than by any universal superiority of non-linear models.

Given this evidence, the usefulness of deep sequence models with many layers is unclear: their complexity, latency, and overfitting risk may offer little advantage over well-regularized linear models with the right features and multi-scale aggregates, or over boosted trees where dense-regime non-linearities matter \cite{sirignano2019universal}. Future research should instead explore lightweight designs that retain interpretability and low latency, and should report the signal share and aggregation bound of Section~\ref{sec:theory} alongside any headline $R^2$.

\paragraph{Limitations.}
The Section~5 evidence comes from a single hour (14:00--15:00 ET) of a single trading day (July 30, 2025) on a single venue, and the reported $R^2$ at longer horizons benefits mechanically from target aggregation; because the underlying commercial data cannot be redistributed, those numbers are not reproducible from the replication repository and should be read as author-reported. The public-data study addresses reproducibility, persistence baselines, statistical significance, definitional sensitivity, and the aggregation mechanism, but covers a single symbol-day in a different market regime and vintage. Neither sample yet evaluates economic value in an execution setting. Extending both samples to multiple days and regimes (including scheduled macroeconomic announcements), benchmarking against order-flow toxicity measures \cite{easley2012vpin}, reframing extreme-withdrawal prediction as tail-event classification, and quantifying execution-cost savings from acting on the forecasts are the natural next steps.

\clearpage

% \clearpage

% \FloatBarrier
% \clearpage

\FloatBarrier

% \section{References}

\appendix

\section*{Appendix A: Raw MBO Data Fields}
\label{app:datafields}

The raw Market-by-Order (MBO) dataset from Databento contains the following fields:

\begin{table}[H]
\centering
\begin{tabular}{p{0.25\linewidth} p{0.7\linewidth}}
\hline
\textbf{Field} & \textbf{Description} \\
\hline
\texttt{ts\_recv} & Capture-server-received timestamp, in nanoseconds since UNIX epoch. \\
\texttt{ts\_event} & Matching-engine-received timestamp, in nanoseconds since UNIX epoch. \\
\texttt{ts\_in\_delta} & Matching-engine-sending timestamp expressed as nanoseconds before \texttt{ts\_recv}. \\
\texttt{channel\_id} & Channel ID assigned by Databento (incrementing integer starting at zero). \\
\texttt{publisher\_id} & Publisher ID assigned by Databento, denoting dataset and venue. \\
\texttt{instrument\_id} & Numeric instrument identifier. \\
\texttt{symbol} & Symbol of the instrument. \\
\texttt{sequence} & Message sequence number assigned at the venue. \\
\texttt{order\_id} & Unique order ID assigned at the venue. \\
\texttt{price} & Order price, stored as an integer where one unit equals $10^{-9}$ (i.e., \$0.000000001). \\
\texttt{size} & Order quantity. \\
\texttt{side} & Side of the event: Bid (buy), Ask (sell), or None if unspecified. \\
\texttt{action} & Event type: Add, Cancel, Modify, Clear Book, Trade, Fill, or None. \\
\texttt{rtype} & Record type (each schema corresponds to one value). \\
\texttt{flags} & Bit field indicating event end, message characteristics, and data quality. \\
\hline
\end{tabular}
\caption{Raw Market-by-Order (MBO) data fields as provided by Databento.}
\end{table}

\section*{Appendix B: Proof of Proposition~\ref{prop:aggregation}}
\label{app:proof}

Let $\mathcal{F}_t$ denote the $\sigma$-field generated by $\{(s_\tau, u_\tau)\}_{\tau \le t}$. Because $u$ is white noise independent of the signal, $\mathbb{E}[y_{t+j} \mid \mathcal{F}_t] = \mathbb{E}[s_{t+j} \mid \mathcal{F}_t] = \varphi^j s_t$ for $j \ge 1$, hence
\[
\mathbb{E}\big[\bar y_{t,k} \mid \mathcal{F}_t\big] \;=\; c_k\, s_t,
\qquad
c_k \;=\; \frac{1}{k}\sum_{j=1}^{k} \varphi^j \;=\; \frac{\varphi\,(1-\varphi^k)}{k\,(1-\varphi)} .
\]
The variance explained by this conditional mean is $c_k^2 \sigma_s^2 = B_k$.

For the target variance, the signal autocovariance is $\gamma_s(d) = \sigma_s^2 \varphi^{d}$, so
\[
\operatorname{Var}\Big(\frac{1}{k}\sum_{j=1}^{k} s_{t+j}\Big)
= \frac{1}{k^2}\Big( k\,\gamma_s(0) + 2\sum_{d=1}^{k-1} (k-d)\,\gamma_s(d) \Big)
= A_k ,
\]
and the independent noise average contributes $\sigma_u^2 / k$, giving $\operatorname{Var}(\bar y_{t,k}) = A_k + \sigma_u^2/k$.

The conditional mean is the $L^2$-optimal forecast given $\mathcal{F}_t$, so the population $R^2$ of the optimal time-$t$ forecast is
\[
\bar R^2(k) \;=\; \frac{\operatorname{Var}\big(\mathbb{E}[\bar y_{t,k} \mid \mathcal{F}_t]\big)}{\operatorname{Var}(\bar y_{t,k})}
\;=\; \frac{B_k}{A_k + \sigma_u^2/k} .
\]
Any forecast measurable with respect to the coarser information set $\sigma(\{y_\tau\}_{\tau \le t}) \subseteq \mathcal{F}_t$ is an $L^2$ projection onto a subspace of $L^2(\mathcal{F}_t)$, so its explained variance---and hence its population $R^2$---is weakly smaller. For $k=1$ the expression reduces to the familiar $\bar R^2(1) = \varphi^2 \sigma_s^2 / (\sigma_s^2 + \sigma_u^2)$. \hfill$\square$

\end{document}